# Controlled exchange interaction between pairs of neutral atoms in an optical lattice

Marco Anderlini*, Patricia J. Lee, Benjamin L. Brown, Jennifer Sebby-Strabley†, William D. Phillips, and J. V. Porto

*Joint Quantum Institute, National Institute of Standards and Technology and University of Maryland, Gaithersburg, MD, 20899, USA*

*Present Address: INFN sezione di Firenze, Via Sansone 1, I-50019 Sesto Fiorentino (FI), Italy

†Present Address: Honeywell Aerospace, 12001 State Highway 55, Plymouth, MN, 55441, USA

**Ultra-cold atoms trapped by light, with their robust quantum coherence and controllability, provide an attractive system for quantum information processing and for simulation of complex problems in condensed matter physics. Many quantum information processing schemes require that individual qubits be manipulated and deterministically entangled with one another, a process that would typically be accomplished by controlled, state-dependent, coherent interactions among qubits. Recent experiments have made progress toward this goal by demonstrating entanglement among an ensemble of atoms[1] confined in an optical lattice. Until now, however, there has been no demonstration of a key operation: controlled entanglement between atoms in isolated pairs. We have used an optical lattice of double-well potentials[2,3] to isolate and manipulate arrays of paired atoms, inducing controlled entangling interactions within each pair. Our experiment is the first realization of proposals to use controlled exchange coupling[4] in a system of neutral atoms[5]. Although $^{87}$Rb atoms have nearly state-independent interactions, when we force two atoms into the same physical location, the wavefunction exchange symmetry of these identical bosons leads to state-dependent dynamics. We observe repeated interchange of spin between atoms**

**occupying different vibrational levels, with a coherence time of more than ten milliseconds. This observation represents the first demonstration of the essential component of a quantum SWAP gate in neutral atoms. The "half implementation" of SWAP, the $\sqrt{SWAP}$ gate, is entangling, and together with single qubit rotations form a set of universal gates for quantum computation[4].**

Particle exchange symmetry plays a crucial role in much of condensed matter physics, for example allowing spin-independent, purely electrostatic interactions between electrons to give rise to magnetism by correlating their spins. While such effects have been extensively discussed in the context of fermions, similar exchange effects also apply to bosons, such as $^{87}$Rb, except here the particle wavefunctions are symmetrised rather than anti-symmetrised. Exchange interactions leading to SWAP operations (interchanging the state of two qubits) have been proposed for entangling qubits in condensed matter implementations of quantum computing[4,6], and as a mechanism for single qubit control in coded qubit spaces[7]. More recently exchange-induced entanglement has been proposed for ultracold neutral atoms[5,8]. Other schemes[9,10,11], not involving exchange, have relied on mechanisms that directly depend on the internal (qubit) state, requiring state-dependent motion, interaction, or excitation of the atoms. Exchange interactions have the advantage that they require none of these. Ordinary state-dependent mechanisms often suffer from decoherence because of state-dependent coupling with the environment. Exchange mechanisms can be relatively free of such decoherence. For example, one could choose magnetic field insensitive states as the qubit basis even if those states had no direct spin-dependent interactions.

To illustrate the working scheme of the two-qubit $\sqrt{SWAP}$ gate with bosons, consider a pair of atoms each occupying the single-particle vibrational ground state of two adjacent potential wells, left ($L$) and right ($R$), with spatial wavefunctions $\phi_L(\boldsymbol{x})$ and $\phi_R(\boldsymbol{x})$ (see Fig. 1a). The full, single-atom wavefunction is $|q_v\rangle = \phi_v(\boldsymbol{x})|q\rangle$, where each qubit (specified by its location $v = \{L, R\}$) can be encoded in two internal spin states of an atom as $|q\rangle = a|0\rangle + b|1\rangle$ for amplitudes $a$ and $b$ associated with the qubit states $|0\rangle$ and $|1\rangle$. For our demonstration, $|0\rangle$ and $|1\rangle$ are Zeeman states of $^{87}$Rb atoms, which are in adjacent sites of a double-well potential[2]. Neutral atoms have short range "contact" interactions, and in $^{87}$Rb are nearly spin-independent. To initiate the interaction, we merge the *L* and *R* sites into a single site so that the atoms' spatial probability distributions overlap[12]. During this merger, the trapping potential is carefully adjusted so that the atoms in *L* and *R* are adiabatically transferred to different vibrational states of the single well[3]: $\phi_L(\boldsymbol{x}) \to \phi_e(\boldsymbol{x})$ and $\phi_R(\boldsymbol{x}) \to \phi_g(\boldsymbol{x})$ (see Fig. 1a). Since the two qubits are encoded in identical bosons, the full two-particle wavefunction must be symmetric under particle exchange, e.g., $|q_L, p_R\rangle = \phi_L(\boldsymbol{x}_1)\phi_R(\boldsymbol{x}_2)|q\rangle_1|p\rangle_2 + \phi_R(\boldsymbol{x}_1)\phi_L(\boldsymbol{x}_2)|p\rangle_1|q\rangle_2$, where the two atoms are labelled 1 and 2. (In the merged trap, the subscripts *L* and *R* are replaced by *e* and *g*, respectively.) The symmetrised states $|0_L, 0_R\rangle$, $|0_L, 1_R\rangle, |1_L, 0_R\rangle, |1_L, 1_R\rangle$ represent a convenient computational basis since the identification of the qubit is straight-forward: $|q\rangle$ is always associated with $\phi_L(\boldsymbol{x})$ (or $\phi_e(\boldsymbol{x})$ when merged), while $|p\rangle$ is always associated with $\phi_R(\boldsymbol{x})$ (or $\phi_g(\boldsymbol{x})$). When the atoms interact in the merged trap, the symmetrised energy eigenstates are no longer the computational basis. The eigenstates are separable into spin and spatial components:

$$|\psi_S\rangle = \phi_S(\boldsymbol{x}_1, \boldsymbol{x}_2)|S\rangle = \left(|1_e, 0_g\rangle - |0_e, 1_g\rangle\right)/\sqrt{2}$$

$$|\psi_T^0\rangle = \phi_T(\boldsymbol{x}_1, \boldsymbol{x}_2)|T^0\rangle = \left(|1_e, 0_g\rangle + |0_e, 1_g\rangle\right)/\sqrt{2}$$

$$|\psi_T^-\rangle = \phi_T(\boldsymbol{x}_1, \boldsymbol{x}_2)|T^-\rangle = |0_e, 0_g\rangle$$

$$|\psi_T^+\rangle = \phi_T(\boldsymbol{x}_1, \boldsymbol{x}_2)|T^+\rangle = |1_e, 1_g\rangle,$$

where $\phi_S(\boldsymbol{x}_1,\boldsymbol{x}_2)=\phi_e(\boldsymbol{x}_1)\phi_g(\boldsymbol{x}_2)-\phi_g(\boldsymbol{x}_1)\phi_e(\boldsymbol{x}_2)$ and $\phi_T(\boldsymbol{x}_1,\boldsymbol{x}_2)=\phi_e(\boldsymbol{x}_1)\phi_g(\boldsymbol{x}_2)+\phi_g(\boldsymbol{x}_1)\phi_e(\boldsymbol{x}_2)$; $|S\rangle=\left(|1\rangle_1|0\rangle_2-|0\rangle_1|1\rangle_2\right)/\sqrt{2}$, $|T^0\rangle=\left(|1\rangle_1|0\rangle_2+|0\rangle_1|1\rangle_2\right)/\sqrt{2}$, $|T^-\rangle=|0\rangle_1|0\rangle_2$, and $|T^+\rangle=|1\rangle_1|1\rangle_2$. The spatial component of the singlet state $|\psi_S\rangle$ is anti-symmetric under exchange of particles; there is no density overlap between the two particles, giving essentially zero interaction energy for the short range contact interactions between the atoms. On the other hand, the triplet states have an interaction energy $U_{eg}=\left(8\pi\hbar^2 a_s/m\right)\int|\phi_e(\boldsymbol{x})|^2|\phi_g(\boldsymbol{x})|^2 d^3x$, where $a_s$ is the s-wave scattering length and $m$ is the mass of $^{87}$Rb[13]. This energy difference between the "singlet" and the "triplet" states can be viewed as arising from an effective magnetic interaction $\propto \boldsymbol{\sigma}_e\cdot\boldsymbol{\sigma}_g$ between atoms in the ground and excited state, where $\boldsymbol{\sigma}_\nu$ is the Pauli spin operator acting on the qubit basis, for the atom in the vibrational state $\nu=\{e,g\}$. This interaction can give rise to a spin exchange oscillation between the qubit states $|0_e,1_g\rangle$ and $|1_e,0_g\rangle$. If atoms in any of the four states of the computational basis are combined into a single site adiabatically with respect to the lattice vibrational level spacing, but diabatically with respect to $U_{eg}$ (thus projecting onto the interacting eigenstates), they evolve in time as shown in Table 1. At time $T_{sw}\equiv\pi\hbar/U_{eg}$, the internal states associated with $\phi_e(\boldsymbol{x})$ and $\phi_g(\boldsymbol{x})$ are swapped. If the interaction is stopped at $T_{sw}/2$ (e.g., by separating the atoms into the *L* and *R* sites), then the result is an entangling $\sqrt{SWAP}$.

We realized this exchange-mediated SWAP operation using arrays of pairs of $^{87}$Rb atoms in a 3D optical lattice. The lattice consists of a dynamically adjustable 2D lattice of double-wells in the horizontal plane[2,3], and an independent 1D lattice along the vertical direction. By controlling the laser polarization, the unit cell of the 2D lattice can be continuously changed between single-well ($\lambda$-lattice) or double-well ($\lambda/2$-lattice) configurations (see Fig. 1a), where $\lambda$=816 nm. We start with a magnetically trapped Bose Einstein condensate of ≈6.0 × $10^4$ $^{87}$Rb atoms in the $5S_{1/2}$ $\left|F=1, m_F=-1\right\rangle$ magnetic state, and slowly (in 140 ms) turn on the $\lambda/2$-lattice and vertical lattice, reaching depths of 40(2) $E_R$ and 54(3) $E_R$, respectively. ($E_R = \hbar^2 k_R^{\ 2} / 2m = h \times 3.45$ kHz is the photon recoil energy and $k_R = 2\pi / \lambda$ is the photon recoil momentum.) Ideally, the ensemble crosses the Mott insulator transition[14], creating a central core of atoms with unit filling factor[15] in the ground state of the $\lambda/2$-lattice. The magnetic confining fields are then turned off, leaving a homogeneous field $B_0 \approx$ 4.85 mT, which defines the quantization axis. It also provides a quadratic Zeeman shift large enough that we can selectively RF couple only the $\left|F=1, m_F=-1\right\rangle$ and $\left|F=1, m_F=0\right\rangle$ states[12], designated as our qubit states $\left|1\right\rangle$ and $\left|0\right\rangle$, respectively. Following this loading procedure, isolated pairs of qubits are in the state $\left|1_L, 1_R\right\rangle$ inside separate unit cells of the lattice (see Fig. 1a, step 1).

We can prepare every pair of atoms in any non-entangled two-qubit state by selectively addressing the atoms in the *L* and *R* sites. We exploit the spin-dependence of the potential, which can be manipulated through the same polarization control used to adjust the lattice topology[2,12]. We first induce a state-dependence in the optical potential that produces an effective magnetic field gradient between the two adjacent sites of the double well. This introduces a differential shift $\Delta\nu_{\mathrm{RF}}$ in the spin-resonant frequencies between the two sites. The *L* or *R* qubits are then selectively addressed by applying a RF pulse resonant only with those qubits. In our experiment, $\Delta\nu_{\mathrm{RF}} \approx$20 kHz and we can prepare the state $\left|0_L, 1_R\right\rangle$ with 95% fidelity.

To measure the qubit state after the double well is transformed into a single well, we map the quasi-momentum of atoms occupying different vibrational bands of the optical potential onto real momenta lying within different Brillouin zones[16,17]. This is achieved by switching off the λ-lattice and the vertical lattice in 500 μs; after a 13 ms time-of-flight, atoms occupying different vibrational levels become spatially separated and can be absorption imaged. Moreover, applying a magnetic field gradient during time-of-flight separates atoms in different spin states along another axis. The populations of atoms in $|0\rangle$, $|1\rangle$ and $\phi_e(\boldsymbol{x})$, $\phi_g(\boldsymbol{x})$ can thus be differentiated in a single image (see Fig. 2). By measuring the population in the different Brillouin zones resulting from the samples loaded either only in the left or only in the right sites of the double wells, we found that more than 80% (85%) of the atoms starting in the *L* (*R*) sites end in the first excited (ground) state of the single-well potential.

As a demonstration of exchange-induced SWAPping, we initially prepare the atoms in the state $|0_L, 1_R\rangle$. We then merge each double well into a single well, transferring the atoms from the *L* and *R* sites into the first excited and ground states, respectively, of the single-well potential. The lattice parameters are adjusted throughout the transformation so that the vibrational frequencies along all three spatial directions remain non-degenerate to avoid unwanted energy level crossings; the lowest vibrational frequency is always along the direction of the double wells. This transformation takes 500 μs, a timescale chosen to be adiabatic with respect to vibration. Since the basis change due to interactions occurs during a small fraction of the total merge time (as indicated by the colour transition at ≈0.45 ms in Fig. 1b), this transformation is nearly diabatic with respect to interactions. This projects the atoms onto a superposition of the two eigenstates $|\psi_S\rangle$ and $|\psi_T^0\rangle$ (see Fig. 1b), which oscillates between the states $|0_e, 1_g\rangle$ and $|1_e, 0_g\rangle$. We calculate that, assuming vibrational adiabaticity, the failure to be completely diabatic would result in approximately 92% population oscillation. (We estimate that the time scale to be fully adiabatic with respect to interactions is larger than 4 ms.) The state evolves in this single well configuration for a hold time $t_h$ before measurement. As shown in Fig. 3, the population in each spin component oscillates between the ground and the first excited states. Fitting an exponentially damped sinusoid to the time-dependent populations in $|0\rangle$ and $|1\rangle$ in the excited state gives a period $2T_{\mathrm{SW}}$ = 285(1) μs, an amplitude of 27(2)%, and a 1/e decay time longer than 10 ms.

The > 10 ms decay of the swap oscillations in Fig. 3 is much longer than the single-spin phase coherence time (≈150 μs)[12]. This long decay time results from the Zeeman-degeneracy of the $|0_e, 1_g\rangle$ and $|1_e, 0_g\rangle$ states, since superpositions of these two-atom states are insensitive to spatial and temporal magnetic field noise, and they form a decoherence free subspace (DFS)[18]. This is similar to fermionic double quantum dot systems[19], but there the underlying noise arises from the inherent fluctuating background of nuclear spins. In contrast, here the inhomogeneous broadening arises from technical sources such as background magnetic field gradients and shot-to-shot field fluctuations. One could choose to encode a single qubit in this two-atom DFS, for which spin exchange would act as a single qubit operation[7]. Here, however, we have sufficient coherence and individual control of the two spins to use the two qubits separately; in this case spin exchange acts to entangle the two qubits.

To investigate spin coherence during the exchange interaction within the full two-qubit Hilbert space, we place both qubits in a superposition of $|0\rangle$ and $|1\rangle$ and allow them to evolve under exchange (see Fig. 4a). Starting with atoms in $|0_e, 1_g\rangle$, we apply an RF π/2 pulse to both qubits, producing a superposition of all four two-qubit logical states. The atoms evolve for 165 μs, a time longer than required for a full swap, and a second π/2 pulse is applied to read out the coherence. (A π-pulse inserted between the π/2 pulses creates a spin echo to cancel the effects of the magnetic field inhomogeneity[12].) The subsequent swap oscillations (Fig. 4c) have the expected phase and 80(2)% of the amplitude compared to the case without the additional RF pulses (Fig. 4b), a degradation approximately consistent with the measured single-qubit decoherence. This shows that the coherence time of the system is longer than the time for both a swap operation and single-qubit operations using RF addressing, which together constitute a set of universal quantum logic operations.

Although the exchange oscillations shows almost no decay over many cycles, the initial amplitude is only 27% of the ideal case. Assuming, pessimistically, that the remaining 73% of the atoms do not SWAP, and so project onto the target state after $\sqrt{SWAP}$ with 50% probability, we find a fidelity of 0.64. The true fidelity is probably higher and can be improved: We believe the major reduction in oscillation amplitude is due to imperfect loading of the initial λ/2-lattice Mott insulator state. Previous experiments in this apparatus[20] indicate that in the λ/2-lattice there are relatively few doubly occupied sites, but there may be a significant fraction of empty sites. An empty site merged with an occupied site produces a site where no SWAPping can occur, reducing the oscillation amplitude. Based on our previous measurements, we estimate approximately 50% of the λ-sites (33% of the atoms) are unpaired. However, this initialization infidelity is distinct from gate fidelity and can be improved[21]. Imperfection in vibrational adiabaticity of the transfer from *L* and *R* to *e* and *g* results in unwanted excitations of atoms to other vibrational states, which are visible in the Brillouin zone mapping of Fig. 2. Such motional problems are likely to be among the limiting factors for the fidelity and speed of any collision-based gate, and will be a topic of future study. Possible improvements include using deeper lattices and coherent control techniques[22]. Imperfections in the RF spin-flip state preparation, the vibrational adiabaticity of the transfer from *L* and *R* to *e* and *g*, and the diabaticity with respect to interactions during the merge account for an amplitude reduction to approximately 59%. Other effects, including the state-dependence of the λ-lattice and of the interaction energies are relatively small. Finally, the coherence of the individual qubits can be significantly improved by actively stabilizing the magnetic field and improving its spatial homogeneity. With the freedom to choose the qubit spin states, we can improve the coherence even further by storing the qubit information in field-insensitive hyperfine "clock-states." In this configuration, site-selective addressing could still be achieved using two-photon transitions[23] through an intermediate site-dependent Zeeman state.

This demonstration of a controlled two-atom exchange operation is the first realization of the key component of an exchange gate in neutral atoms. As with all ensemble qubit measurements[1], we do not directly show non-classical correlations, but our observed spin SWAPping oscillations are clearly indicative that during every SWAP cycle the system undergoes the entangling/disentangling dynamics associated with a $\sqrt{SWAP}$ operation. Our results show that the double well optical lattice can be used as a testbed for exploring the two-atom dynamics that underlie some of the key challenges in neutral atom based quantum computing. Scaling to a large number of individually controlled qubits requires individual and pairwise addressing, which could be accomplished with state-dependent focused laser beams[24]. The direct observation of exchange interactions is also relevant for proposals to engineer quantum spin systems[8,25] where tunnelling and exchange give rise to an effective magnetic interaction between nearest neighbours, $\propto \boldsymbol{\sigma}_i \cdot \boldsymbol{\sigma}_{i+1}$. The direct on-site exchange interaction observed here, $\propto \boldsymbol{\sigma}_e \cdot \boldsymbol{\sigma}_g$, could be used to provide effective magnetic interactions between atoms in different vibrational bands[26,27], or to "stroboscopically" generate magnetic interactions between nearest neighbours[28,29].

| **Table 1 Truth Tables for SWAP and $\sqrt{SWAP}$ gates**[1] | | | |
|---|---|---|---|
| Initial | State after time $t$ | $\sqrt{SWAP}$ $t = \pi\hbar / 2U_{eg} = T_{SW} / 2$ | SWAP $t = \pi\hbar / U_{eg} \equiv T_{SW}$ |
| $\left\|0_e,0_g\right\rangle$ | $e^{-iU_{eg}t/2\hbar}\left\|0_e,0_g\right\rangle$ | $e^{-i\pi/4}\left\|0_e,0_g\right\rangle$ | $\left\|0_e,0_g\right\rangle$ |
| $\left\|0_e,1_g\right\rangle$ | $\cos\left(U_{eg}t/2\hbar\right)\left\|0_e,1_g\right\rangle - i\sin\left(U_{eg}t/2\hbar\right)\left\|1_e,0_g\right\rangle$ | $\left(\left\|0_e,1_g\right\rangle - i\left\|1_e,0_g\right\rangle\right)\Big/\sqrt{2}$ | $\left\|1_e,0_g\right\rangle$ |
| $\left\|1_e,0_g\right\rangle$ | $-i\sin\left(U_{eg}t/2\hbar\right)\left\|0_e,1_g\right\rangle + \cos\left(U_{eg}t/2\hbar\right)\left\|1_e,0_g\right\rangle$ | $\left(-i\left\|0_e,1_g\right\rangle + \left\|1_e,0_g\right\rangle\right)\Big/\sqrt{2}$ | $\left\|0_e,1_g\right\rangle$ |
| $\left\|1_e,1_g\right\rangle$ | $e^{-iU_{eg}t/2\hbar}\left\|1_e,1_g\right\rangle$ | $e^{-i\pi/4}\left\|1_e,1_g\right\rangle$ | $\left\|1_e,1_g\right\rangle$ |

[1] Ignoring a global phase factor $e^{-iU_{eg}t/2\hbar}$.

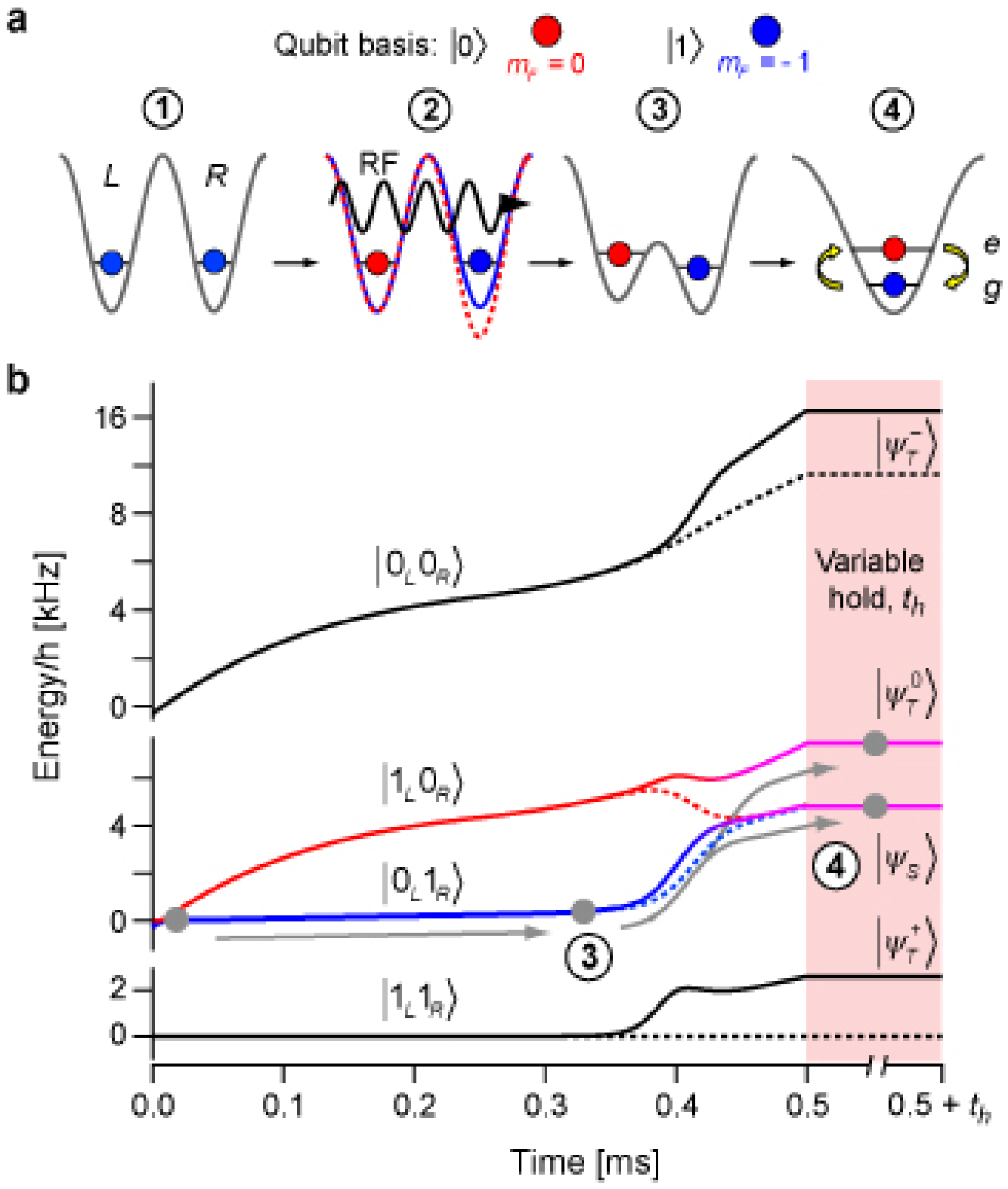

**Figure 1 | Experimental sequence. a,** Preparation and interaction of two qubits. Step 1: the system is initialized as qubit state $|1_L,1_R\rangle$; step 2: the two neighbouring atoms in a double well are prepared in the qubit state $|0_L,1_R\rangle$ using site-selective RF addressing based on the spin-state dependence of the potential (indicated by the differing blue and red potentials); step 3: the potential barrier between the two sites is then lowered; step 4: the two sites merge, allowing the atoms to interact. Careful control of the potentials during this merger forces the atom in the left site into the first excited state and the atom from the right site into the ground state of the final single-well configuration. **b,** Plot of the interacting (solid lines) and non-interacting (dashed lines) two-particle energies during the gate sequence (steps 2 to 4 in part **a**). For visual clarity the energies are relative to the non-interacting $|1_L,1_R\rangle$ eigenenergy, and

the 34 MHz Zeeman shifts are not included. The gray arrows indicate the evolution of the state $\left|0_L,1_R\right\rangle$ from step 2 to step 4. The colour transition from red ($\left|1_L,0_R\right\rangle$) and blue ($\left|0_L,1_R\right\rangle$) to purple ($\left|\psi_T^0\right\rangle$ and $\left|\psi_S\right\rangle$) indicates the mixing of the two logical qubit states. The evolution from the initial state $\left|0_L,1_R\right\rangle$ is non-adiabatic with respect to interactions, and the projection onto the final singlet/triplet eigenstates results in spin exchange oscillations.

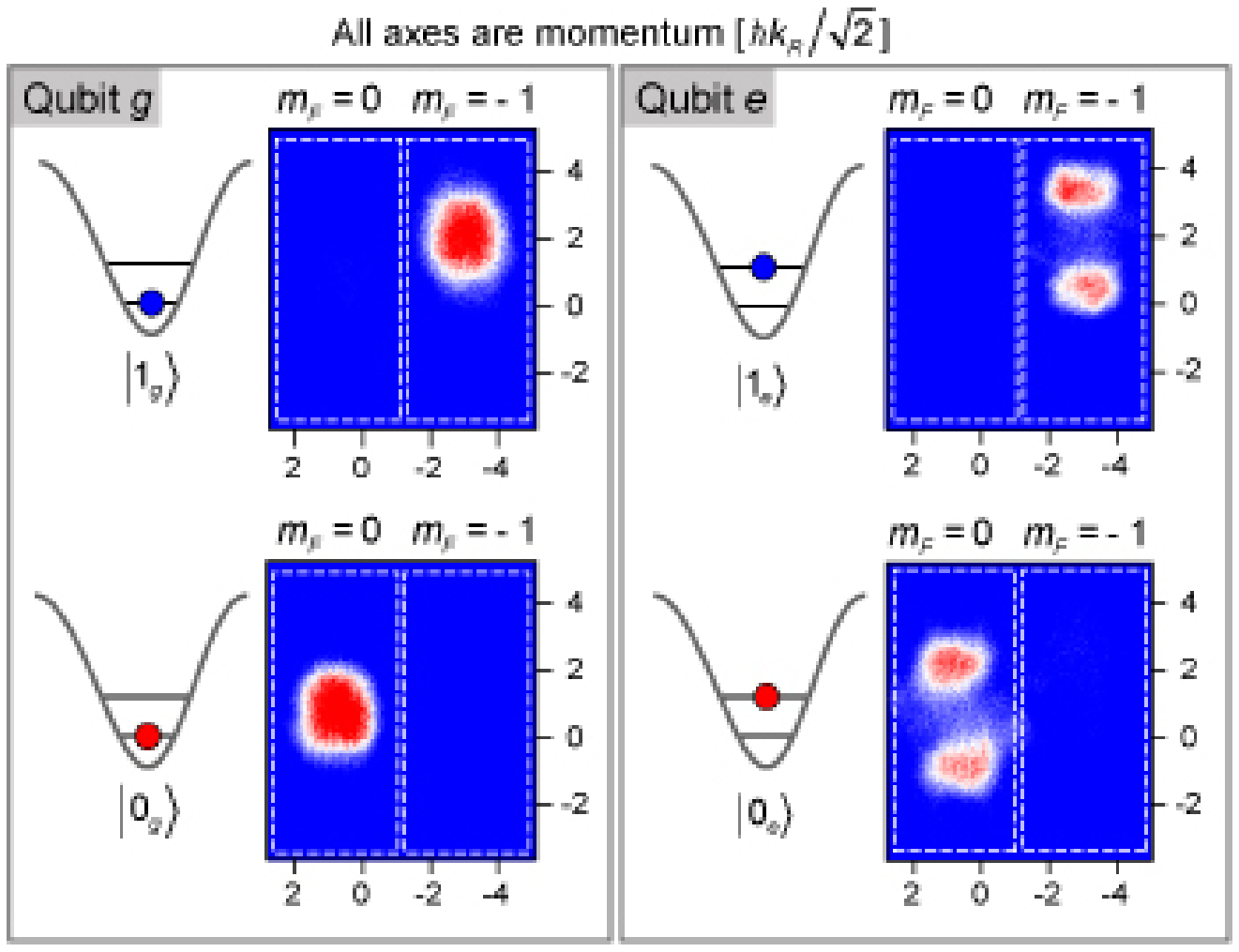

**Figure 2 | Qubit state analysis.** Time-of-flight images mapping the atoms' internal and vibrational states: the images were produced by preparing single atoms in one of the two single-qubit basis states (internal spin states) in either the *L* or *R* qubit and performing the full sequence (steps 2 to 4 in Figure 1 **a**), followed by the Brillouin zone mapping (see text) and time-of-flight absorption imaging. Different vibrational states are thus mapped to different momentum regions. In addition, a magnetic field gradient (diagonal in the image plane) applied during time-of-flight spatially separates atoms in different spin states, indicated by the white dashed-line boxes. Each of the input states maps to a distinct region of the image, allowing us to measure the populations in the spin state $|0\rangle$ or $|1\rangle$ separately for each qubit.

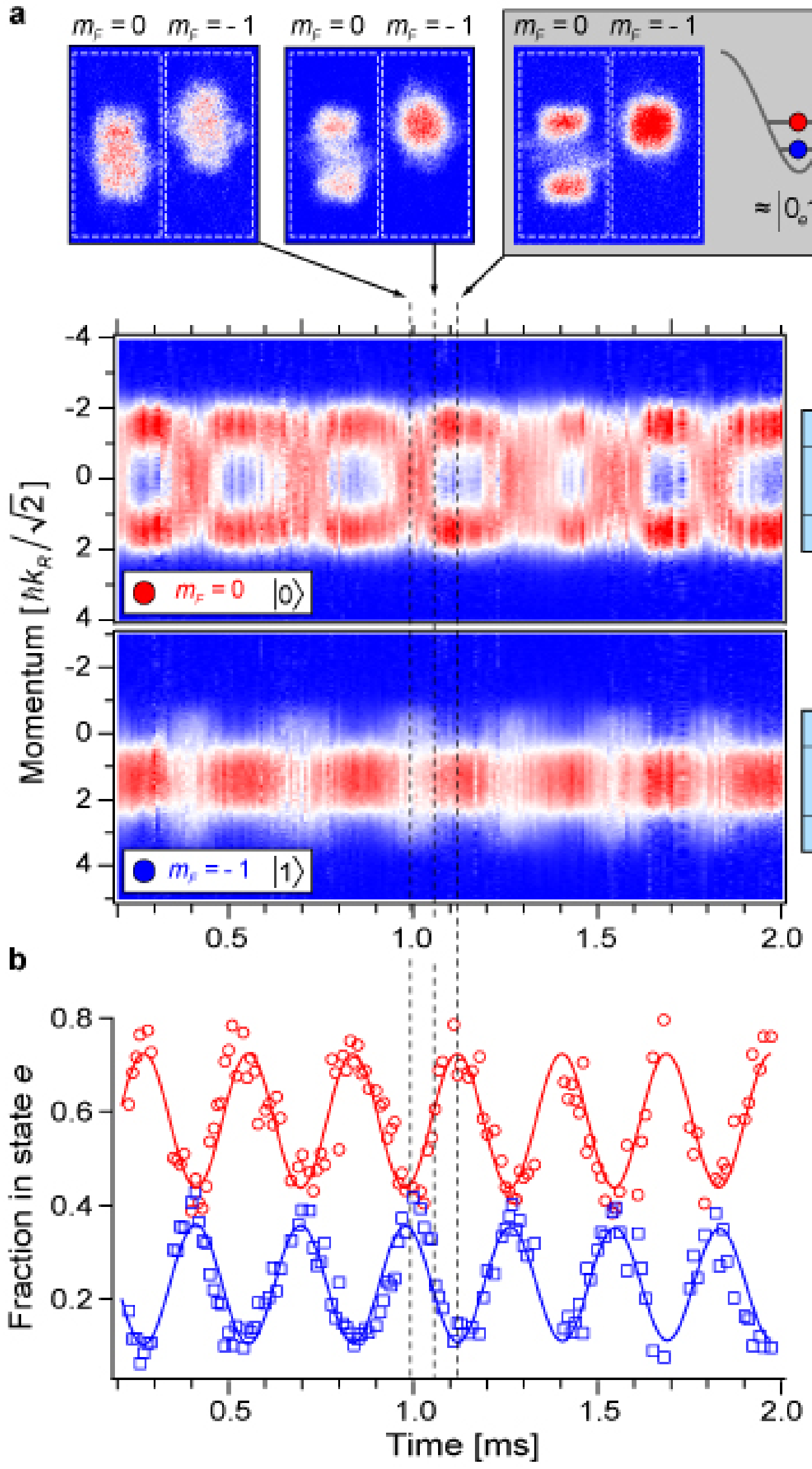

**Figure 3 | Collisional swap dynamics. a,** Concatenated slices of absorption images as a function of hold time $t_h$ in the single-well configuration (Fig. 1a, step 4). For technical reasons, the hold time can be no less than 200μs. Atoms in each vibrational level oscillate between spin states $|0\rangle$ and $|1\rangle$. **b,** Fraction of atom populations in the excited state for atoms in $|0\rangle$ (red) and $|1\rangle$ (blue). Each point is extracted from the data in **a** by fitting the time-of-flight image slices and extracting the relative amount of population in each Brillouin zone. The solid

lines are sinusoidal fits to the data, with a common period of 285(1) μs and a common amplitude of 0.27(2). The amplitude of the oscillation is smaller than the initial excited $|0\rangle$ (or ground $|1\rangle$) fraction, which gives rise to the difference in the bottom two panels of **a** and the offset of the $|0\rangle$ and $|1\rangle$ fractions in **b**.

The phase of the oscillations is affected by interaction during the merging and during the process of switching off the lattice. After more than 6 full periods of oscillation, corresponding to 24 $\sqrt{SWAP}$ cycles, the amplitude of the oscillations shows negligible decay. If the qubits are prepared initially $|1_e,1_g\rangle$ or $|0_e,0_g\rangle$, we observe no evolution of the spin populations.

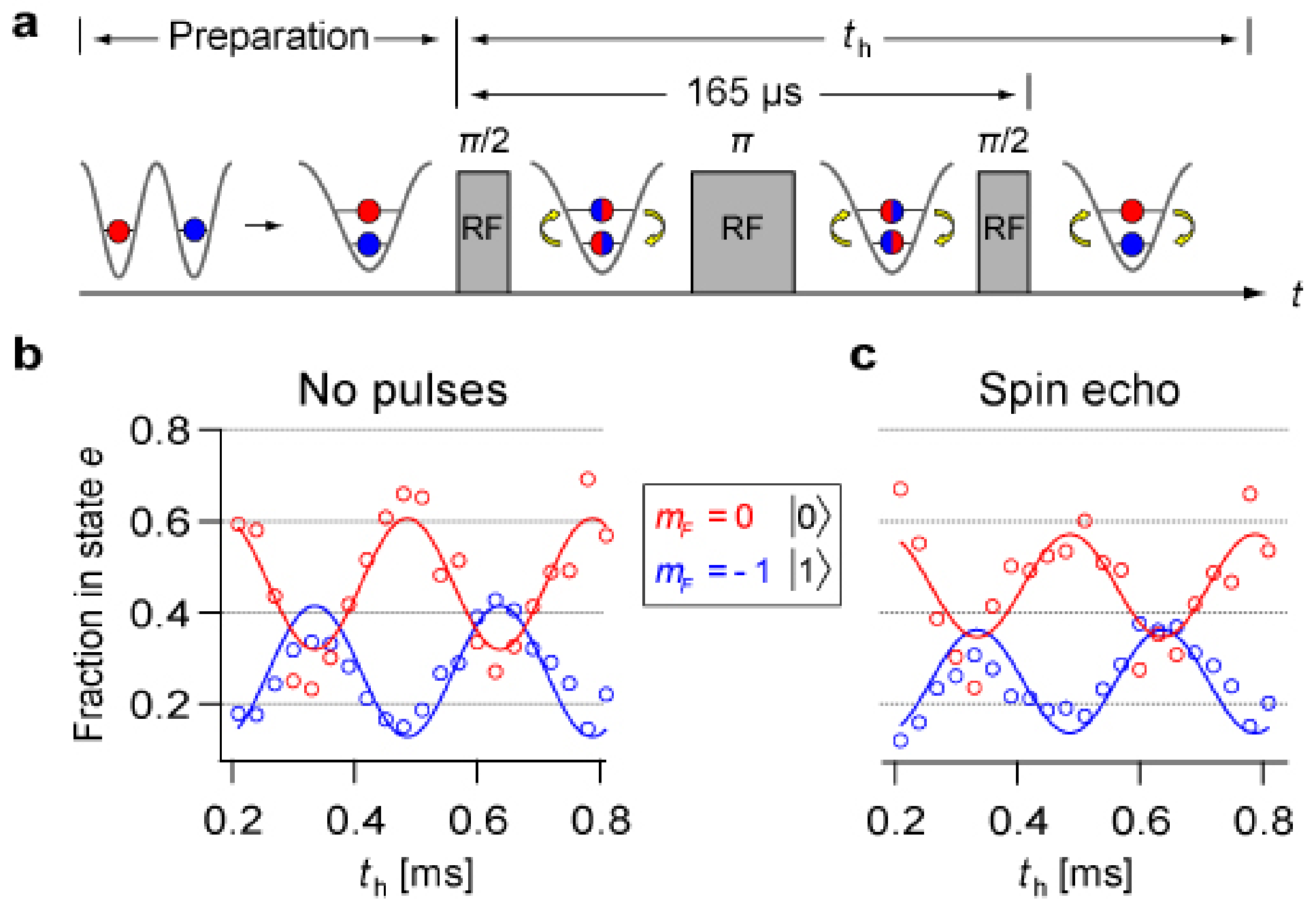

**Figure 4 | Spin-phase coherence during the SWAP operation. a,** The temporal sequence of the experiment. Each pair of atoms in a double-well is initially prepared in the state $|0_L, 1_R\rangle$, and transferred into the single-well configuration. The measured fraction of atom population in the excited state for atoms in $|0\rangle$ (red) and $|1\rangle$ (blue) *vs.* $t_h$ is plotted for **b,** a control case identical to the conditions in Fig. 3, where no additional RF pulses are applied, and **c,** the spin echo case. Between the two $\pi/2$ pulses of the spin echo sequence the atoms were in a superposition of all possible spin states while undergoing a full swap. The exchange oscillations following the spin-echo sequence shown in **c** indicate that spin coherence is preserved during the swap.

**Acknowledgements** The authors thank Ian Spielman and Steve Rolston for contributions to the project, and Ivan Deutsch for helpful discussions. P.J.L., B.L.B., and J.S.-S. acknowledge support from the National Research Council Postdoctoral Research Associateship Program. This work was supported by DTO, ONR, and NASA.

**Author Information** Reprints and permissions information is available at www.nature.com/reprints. The authors declare no competing financial interests. Correspondence and requests for materials should be addressed to J.V.P. (trey@nist.gov).